# HARMONI at ELT: LPOU calibrations for a micron-level pointing accuracy

Gonzalo José Carracedo Carballal [a], Irene Ferro Rodríguez [a], David L King [b], Iago Funes Vecino [a], Guillermo Mercant Rubio [a], Alonso Álvarez Urueña [a], Laura García Moreno [a], Heribert Argelaguet Vilaseca [a], Javier Piqueras López [a], Chiara Cerruti [a]

[a] Centro de Astrobiología (CAB), CSIC-INTA, Ctra. de Ajalvir, km 4, 28850 Torrejon de Ardoz, Madrid, Spain
[b] Centre for Advanced Instrumentation, Lower Mountjoy, South Road, Durham DH1 3LE, U.K.

## ABSTRACT

HARMONI is the first light, adaptive optics assisted, near-IR integral field spectrograph for the ELT [1]. It covers a spectral range from 800 nm to 2450 nm with resolving powers from 3000 to 7000 and spatial sampling of 25 mas and 6mas. It can operate in two adaptive optics modes - SCAO (including a high contrast capability) and MCAO. The project is resuming its final design phase after a rescope design phase in 2025.

To achieve its optimal image quality, HARMONI needs to perform pointing error measurement by means of three guide probes. Each guide probe is a robotic arm with a shoulder-elbow stage that is optically equivalent to two periscopes connected together. Since any systematic positioning error of the shoulder-elbow motors will result in a measurement error of the pointing error itself, an appropriate geometric calibration reference is necessary.

In previous designs, this calibration reference took the form of a deployable calibration mask that could be inserted in the telescope's focal plane. However, this approach prevents the guiding sensors from distinguishing motor contributions from other effects caused by the instrument optics.

In this work, we explore the possibility of inserting the calibration mask as close to the guide probes as possible, along with the addition of the necessary refocus optics in the light path of the guiding sensors. We discuss different variations of this idea, their effect on the image quality, and their ability to resolve positioning errors smaller than 1 µm. The results of these analyses were cross-validated by means of optical simulations in RayZaler, and will be used to put bounds to the expected calibration accuracy achievable by this design.

## 1. INTRODUCTION

HARMONI will be one of the first-generation instruments for the ELT, and it will be used as a “workhorse instrument” for scientific observations. HARMONI is designed to operate in the 0.45µm-2.45µm range (3500<R<18000) in a broad variety of scientific programs. Its different spectral set-ups will enable observations in the NIR bands, including J (1.05µm-1.32µm), H (1.44µm-1.82µm), K (1.95µm-2.47µm) and portions thereof. It will enable both AO (including multi-conjugate AO —MCAO—, single-conjugate adaptive optics —SCAO—, and high-contrast adaptive optics —HCAO—) and non-AO observations, with fields of view (FoV) ranging from 5.3 arcsec × 3.8 arcsec to 1.3 arcsec × 0.9 arcsec.

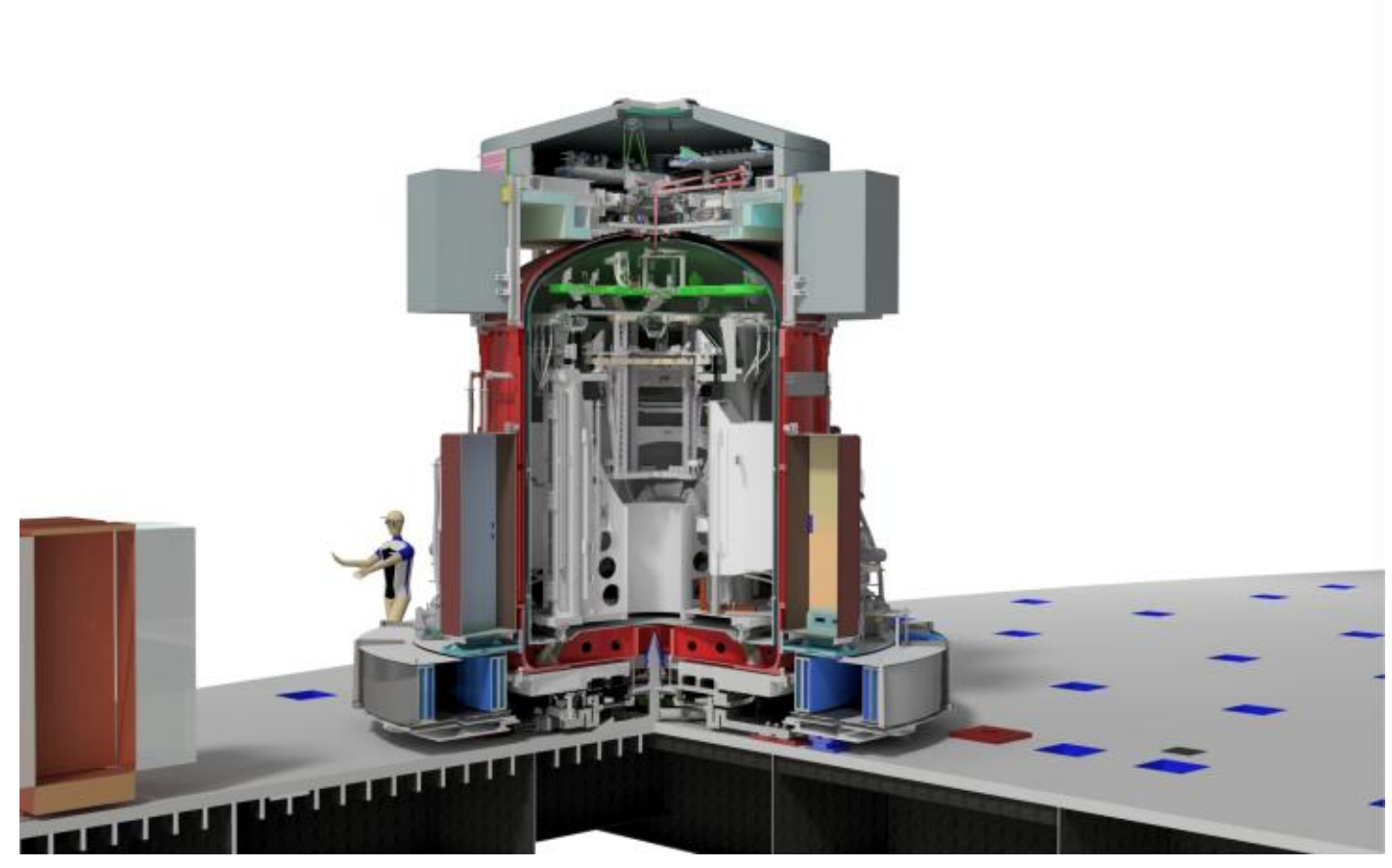

Figure 1: Full section of HARMONI. In red: HARMONI's rotating cryostat.

In its highest resolution configuration, it will be able to resolve objects at the diffraction limit of the ELT, with a sampling of 6 mas/spaxel over a 1.3 arcsec × 0.9 arcsec FoV. In this mode, both HARMONI and ELT must co-operate in closed loop mode, correcting its pointing continuously (order ~ 1 Hz) at the same time it performs WFE sensing and acquires science data. This is done by measuring the position of natural guide stars (NGS) in the surrounding area around the science field (also known as technical field). The contribution of Center for Astrobiology (CAB, CSIC-INTA) to HARMONI is focused on two subsystems: the Calibration Module (CM), responsible for scientific calibrations of the instrument and the Low Order Wavefront Sensing Subsystem (LOWFS) [2] which contains, among other components, the guide probes used for automatic guiding.

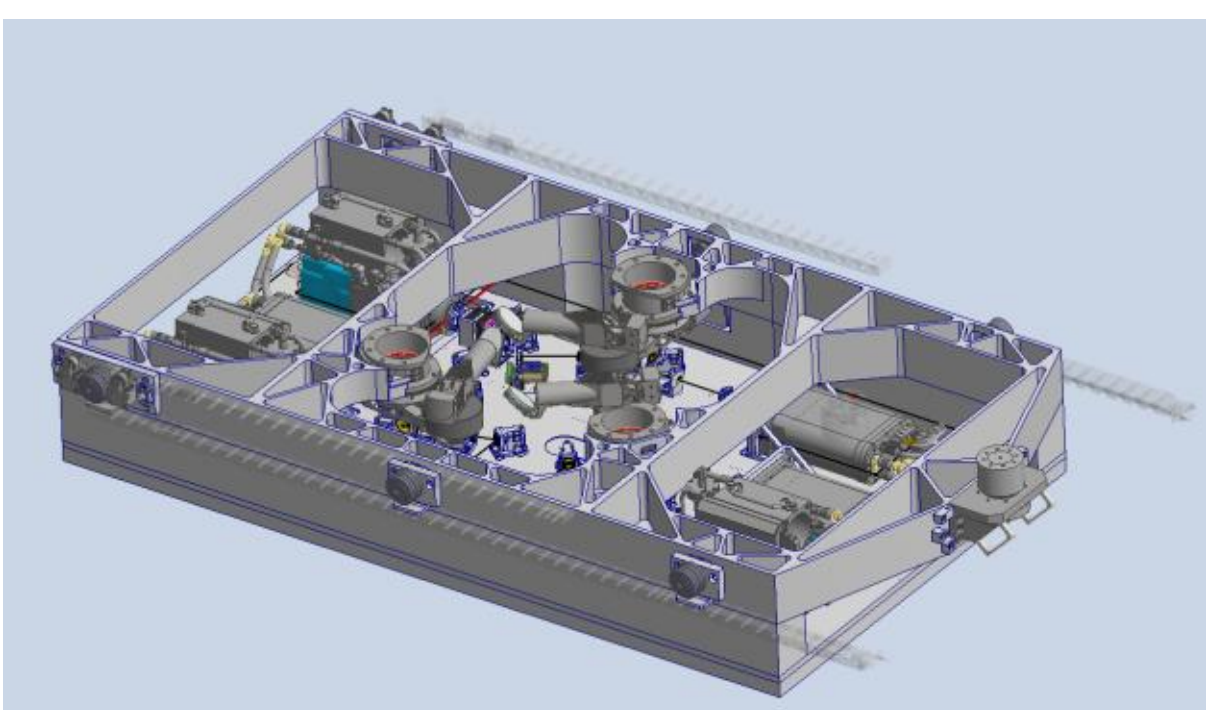

Figure 2: LOWFS, featuring its two sub-units. On top, the LOWFS Pick-Off arm unit (LPOU), supporting the three guide probes. Below, the Low Order Bench (LOB) containing the rest of the optical path of each guide probe.

In the recent years, and in the context of a PhD thesis, CAB developed a free opto-mechanical simulation software named RayZaler with the aim of supporting the different analyses performed during the design phase of LOWFS and the CM. RayZaler has been released to the public in SPIE Astronomical Telescopes and Instrumentation [3], and it is currently being used in a systematic way at CAB as part of its assigned working packages.

In this work, we illustrate the capabilities of RayZaler in a fictional (but more realistic) scenario: what if HARMONI's guide probes required calibration? We will depart from an existing Zemax model of the light path of one of HARMONI's guide probes, and attempt to modify it to support an additional feature. The results of this analyses will be cross-validated with Zemax, demontrating its alignment with existing simulation software.

Begin the Introduction two lines below the Keywords. The manuscript should not have headers, footers, or page numbers. It should be in a one-column format. References should be listed at the end of the manuscript and numbered consecutively in the order of their citation in the text. In-text citations can use superscript or bracketed reference numbers.

## 2. PROBLEM STATEMENT

The guide probes are 3 robotic arms called LOWFS pick-off arms (LPOA). They can be found inside HARMONI's Low Order Wavefront Sensing Subsystem (LOWFS). Each guide probe can be seen as two attached, articulated periscopes that can extract the light from individual guide stars. The light is delivered to the Low Order Bench (LOB), which contains the guiding detectors and most of LOWFS optical components. By measuring the distance of the imaged star between its espected location, LOWFS can be used to deduce the pointing error of the ELT.

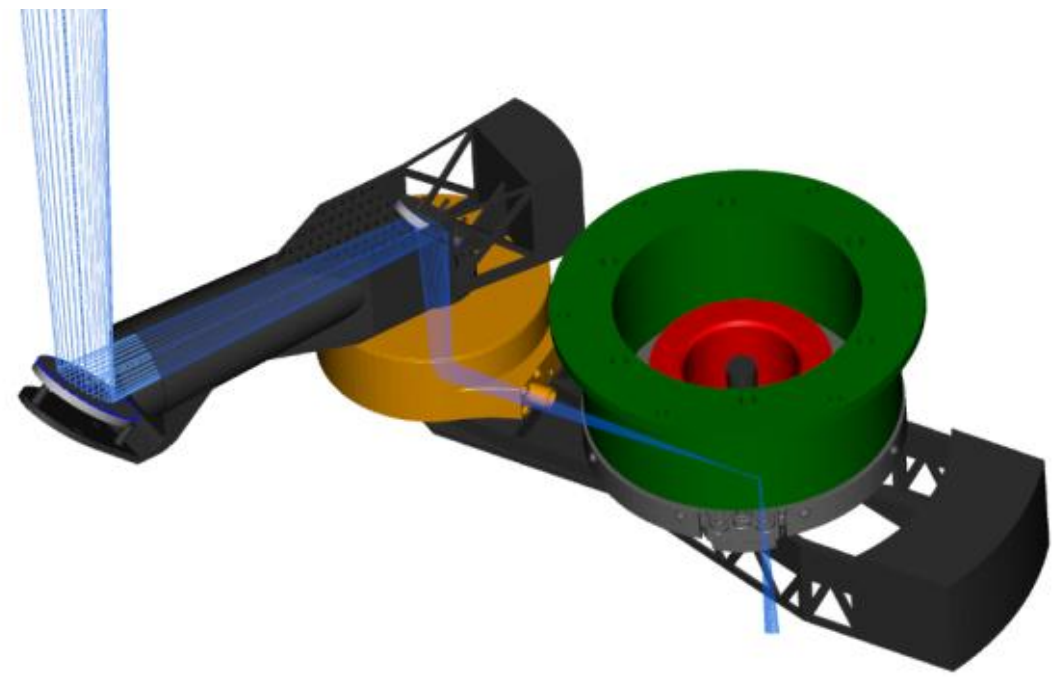

Figure 3: Detail of a LPOA, one of HARMONI's guide probes inside LOWFS. In blue: the light beam of a guide star observed by the guide probe.

In this hypothetical scenario, we impose that the guide probes are affected by systematic effects that cannot be known before LOWFS is installed in the instrument. This would require an on-site calibration reference. One may consider relying on the geometric calibration sources installed before the relay optics provided by MORFEO. However, if observed from the guide probes, the measurement of these calibration sources would be affected by contributions *outside* the guide probes. In particular, they would be affected by optical distortions caused by MORFEO's relay optics and the current rotation of HARMONI's derotator. It would be therefore more interesting to deploy this calibration source between MORFEO and HARMONI, independent from the rotation of the derotator and any optics outside LOWFS.

## 3. PROPOSED SOLUTION

The proposed solution consists of a calibration mask with a certain pattern of point sources that can be deployed between MORFEO and HARMONI. Since there are no focal plane between both instruments, the calibration mask will be inevitably out of focus, and an additional focus optics will be necessary. The proposed focus optics consists of an electronically deployable lens in the optical path of each guide probe. Since the locations where this lens can be placed are restricted by the placement of other components inside of LOWFS, we should be prepared for a suboptimal magnification ratio (i.e. less than unity).

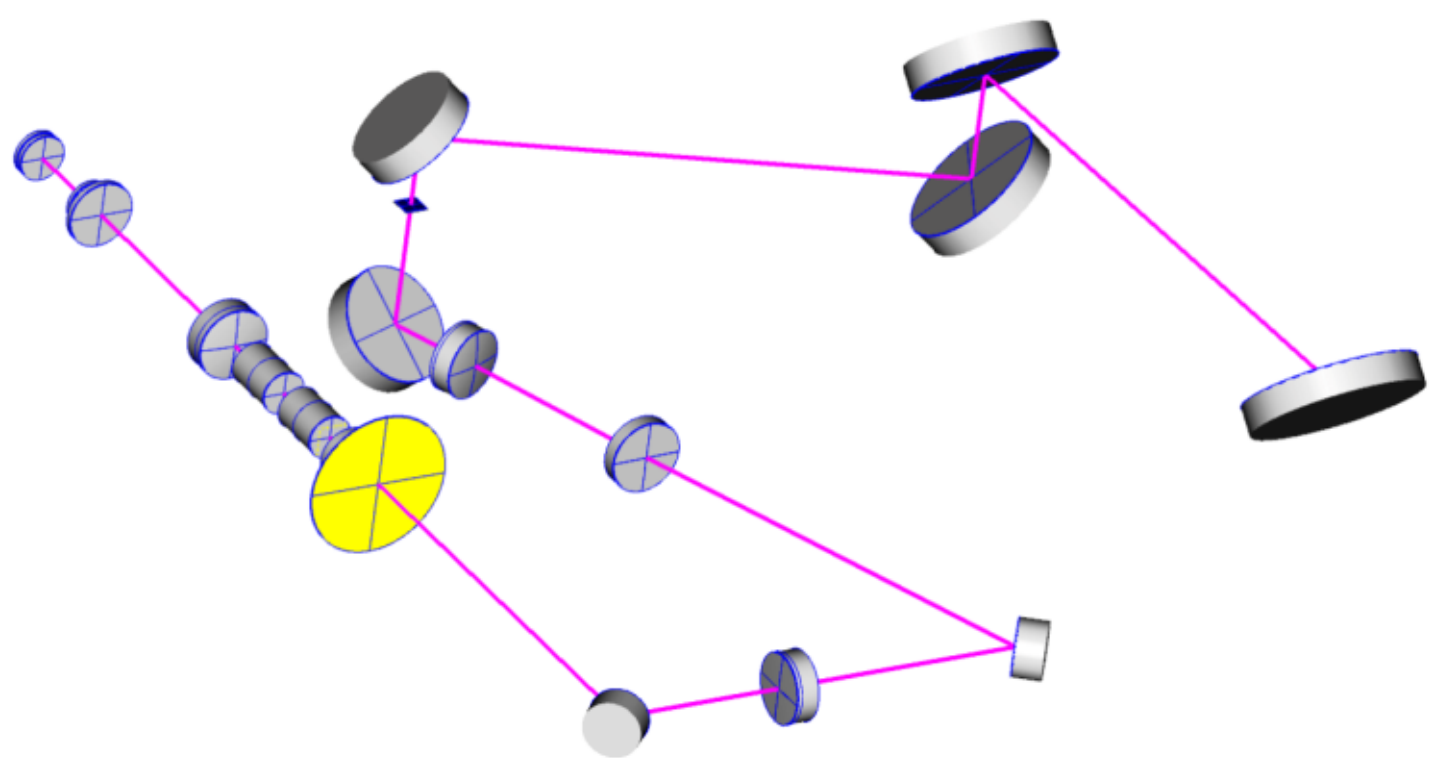

Figure 4: Placement of the calibration lens in LOWFS, highlighted in yellow. The largest 4 mirrors on the top-right corner of the image represent the fold mirrors of the guide probe. The magenta lines represent the sequence of mirrors traversed by light when observing a guide star or a calibration reference.

The chosen location for the calibration mask was chosen immediately above HARMONI's entrance window, 600 mm above the next focal plane inside the instrument. The chosen location for the corresponding deployable lens was between the last fold mirror in the LOB and the first surface of the LOWFS collimator. The acceptance criterion of this solution was conditioned to the following requirements:

- The magnification ratio cannot be less that one order of magnitude below unity (i.e. anything below 1/10 is considered non-acceptable).
- The optical distortion may not affect the linearity of the measurement significantly. Any deviation from linearity must be well below the considered wavelength (1.49 µm).

## 4. METHODS

For these analysis, an optical model of one of the guide probes (provided by Dr. David King) was used as the starting point. It was later modified to incorporate the proposed calibration lens. The RayZaler model file was generated by a Python script that processed the Zemax model's optical prescription to produce an equivalent RayZaler model file.

The generated RayZaler model was cross-validated with Zemax in an unmodified version of the model, in which the light from a star at the zenith illuminated the entirety of ELT's aperture. The acceptance criterion consisted on making sure that the resulting spots in the detector surface were *similar enough* in both simulations. This enough similarity consisted on making sure that the size of the spots differed in less than one wavelength.

Next, both models were modified to incorporate this "calibration lens" at its intended position, and rays were casted from point sources in hypothetical calibration mask with an f/# large enough (60.4) to prevent vignetting. The calibration lens was modelled as a paraxial lens with large aperture. The validation was performed by observing 5 point sources from the LOB detector: One in the center, perfectly aligned with the guide probe position, and 4 at a fixed distance ($\Delta x$ = 4 mm) in the positive and negative directions of the X and Y axes of the calibration mask. The acceptance criterion in this case was that the center of the observed spots must agree within one wavelength.

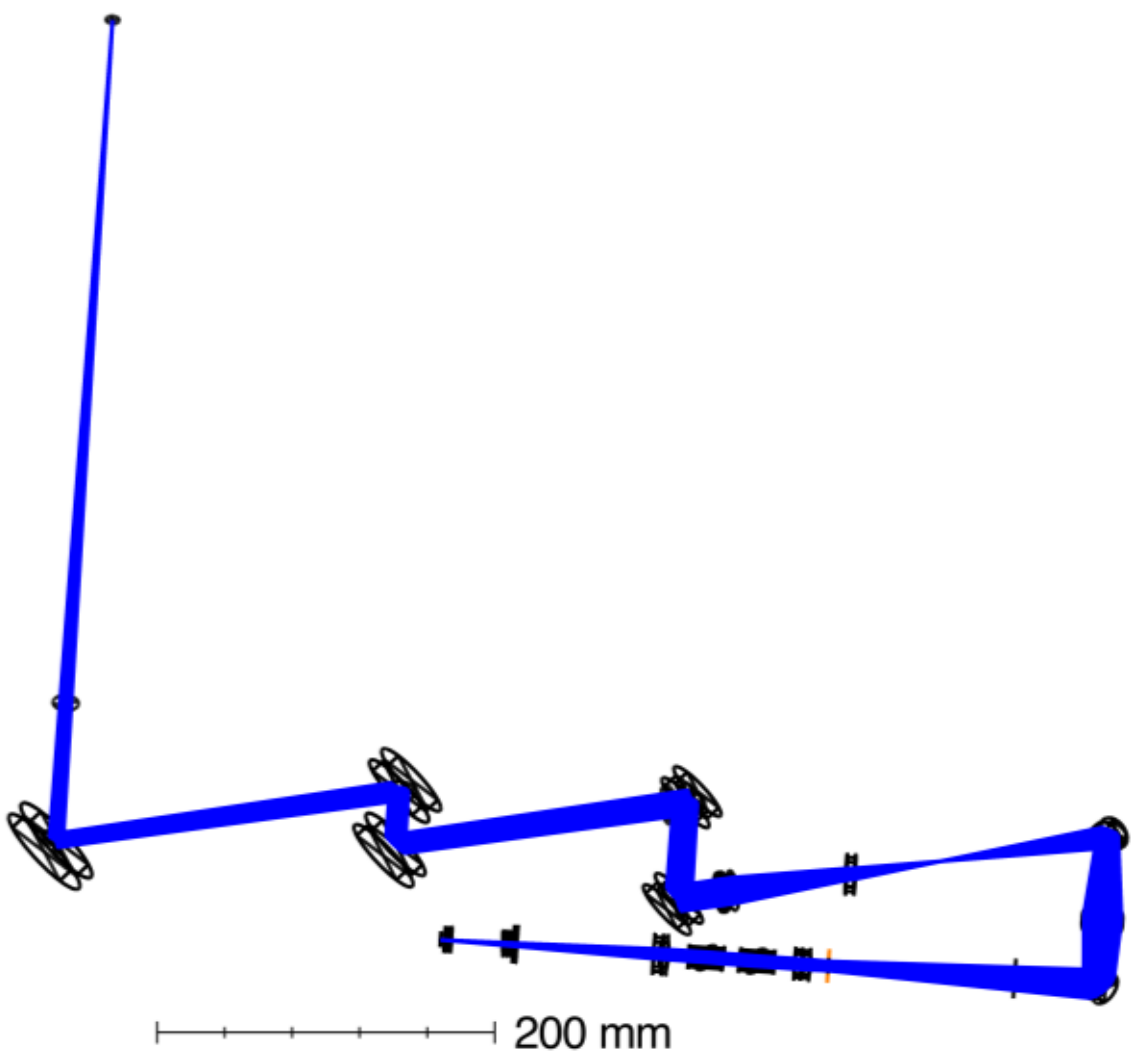

Figure 5: Simulation of LOWFS with the insertion of the calibration lens. Light rays are being casted by a calibration mask 600 mm above the focal plane relayed by MORFEO.

Once the modified model was validated, the central source was displaced in fixed steps of 1 mm along the calibration mask and its observation simulated from RayZaler. The linear fit between the locations of the displaced sources and the locations of the centers of the observed focii was used to determine the final magnification ratio.

## 5. RESULTS

The simulation of the unmodified models with a star at the zenith yielded nearly identical results: the spot centers and widths differed by less than 1 µm. On the other hand, a vertical, candy-shaped caustic was longer in RayZaler. The length of this caustic was observed to be highly sensitive to the wavelength, the beam sampling, and the refractive index of air.

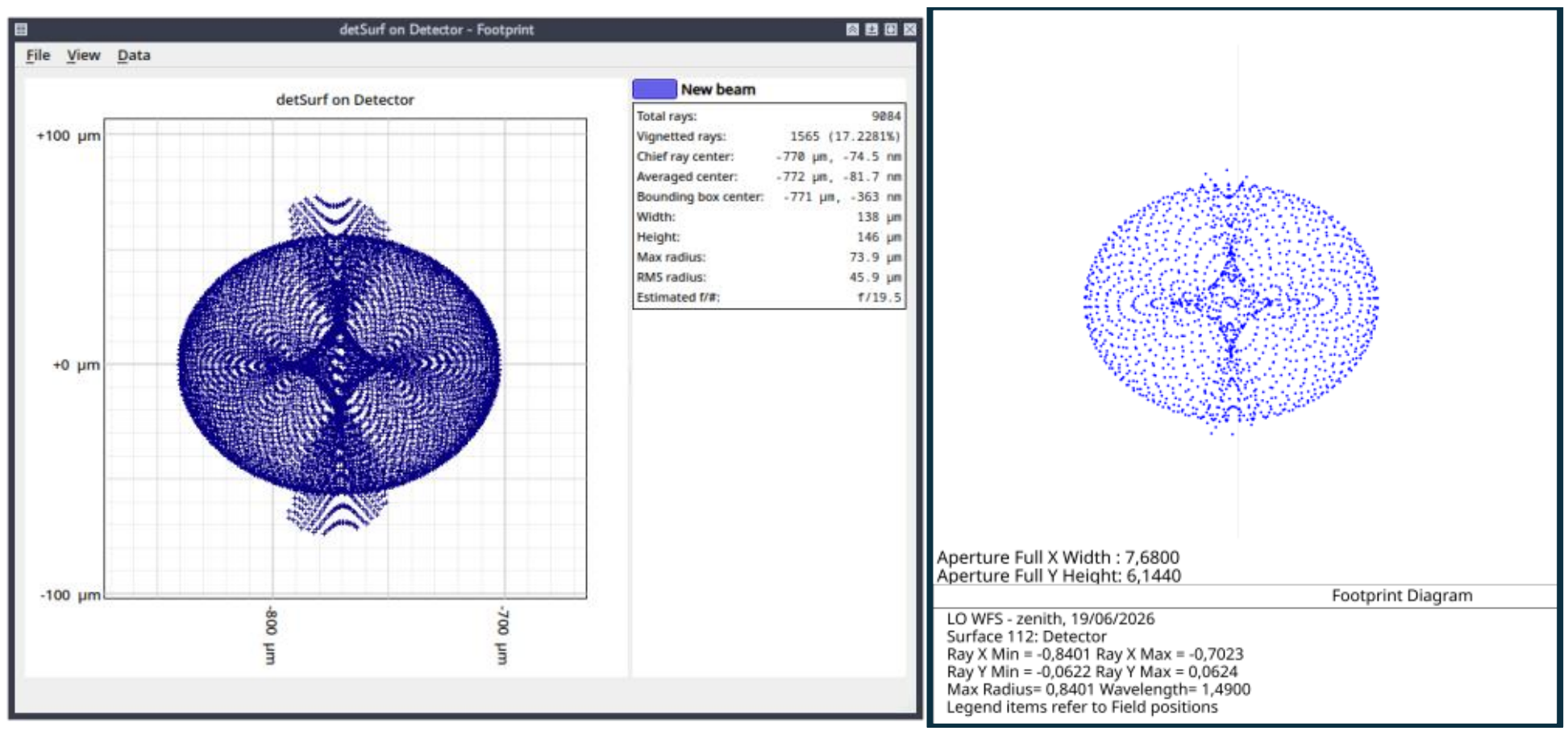

Figure 6: Spot diagram on the LOB's detector, according David King's optical model of LOWFS. The guide probe is actuated to the (0, 0) position. Left: spot diagram from RayZaler. Right: same spot diagram, retrieved from Zemax. The differences were attributed to beam sampling and the numerical precision of the definition of the refractive index of air.

The simulations with the calibration lens were also identical up to differences below 1 µm, mainly attributed to beam sampling and the vertical caustic.

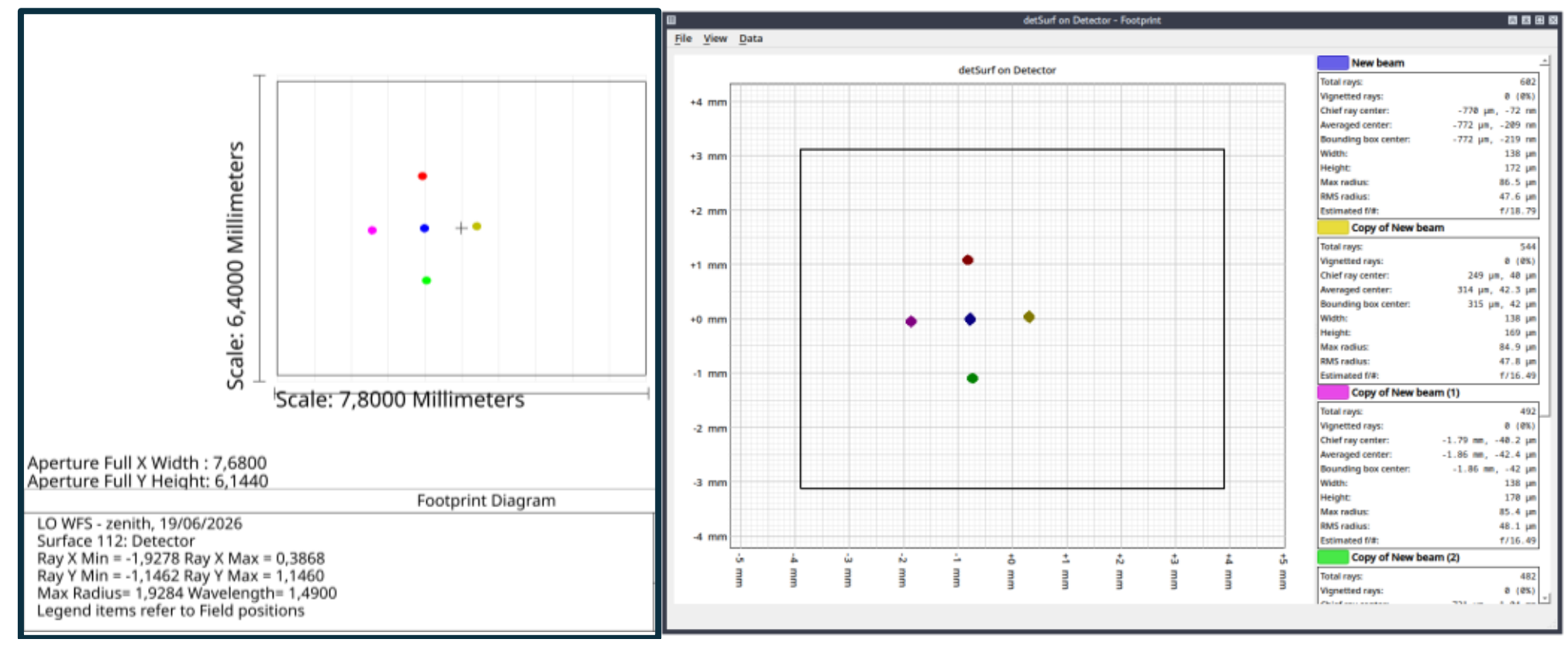

Figure 7: Spot diagram of the imaging of 5 point sources laid out on the hypothetical calibration mask. The point sources were displaced along the positive and negative X and Y axes with respect to the central point source in steps of 4 mm.

With this equivalence demonstrated, the RayZaler model was simulated again with increasing displacements of the source location along the calibration mask, until its beam was vignetted by the optics. The resulting focii were compared to the source locations to deduce the magnification ratio and linearity of the modified system. The resulting magnification ratio was of 1/3.69, with a RMS deviation from linearity of 150 nm dominated by ray sampling.

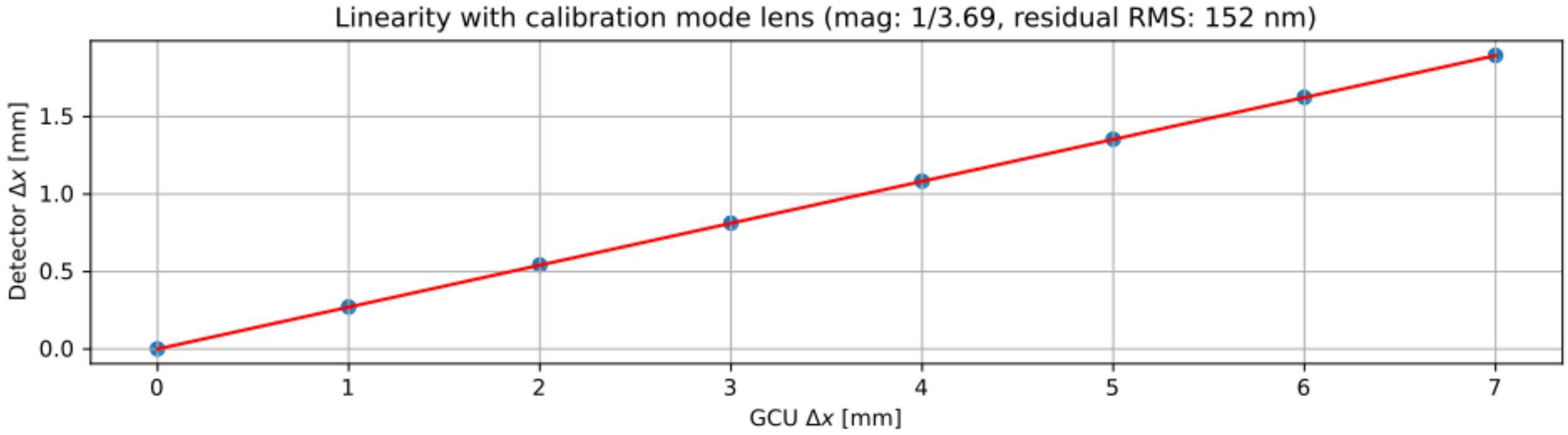

Figure 8: Linear fit between the displacement of the central field in the calibration mask, and the displacement of the observed spot in the LOB's detector. This linear fit was used to deduce the magnification ratio and the deviation from linearity. The deviation of linearity was later attributed to the sampling of the light beam.

## 6. CONCLUSIONS

The RayZaler simulation results showed good agreement with those obtained using Zemax Optic Studio, providing confidence in the validity of the modelling approach. This demonstrates the potential of RayZaler to support the analyses of the performance of optical designs within HARMONI.

## ACKNOWLEDGEMENTS

This work has been supported by grants PID2022-140483NB-C22 and PID2024-158231NB-C22, funded by AEI 10.13039/501100011033 and the European Union, and by grant BDC20221289, funded by the Spanish Ministry of Science and Innovation through the Recovery, Transformation and Resilience Plan, and by NextGenerationEU from the European Union through the Recovery and Resilience Facility.